\documentclass[sigconf,natbib=true,screen=true]{acmart}

\acmSubmissionID{460}

\usepackage{color}
\usepackage{bbm}
\usepackage{multirow}
\usepackage[inline]{enumitem}
\usepackage{graphicx}
\usepackage{subcaption}
\usepackage{sistyle}
\usepackage[ruled]{algorithm2e}
\usepackage{booktabs}
\usepackage{caption}
\SIthousandsep{,}
\usepackage{makecell}
\usepackage{amsmath}
\usepackage[english]{babel}
\usepackage{hyperref}
\usepackage{array} 
\usepackage{graphicx}
\usepackage{algorithmic}
\usepackage{tikz}
\usepackage{wrapfig}
\usepackage{acronym}
\newcommand{\heading}[1]{\vspace*{1mm}\noindent\textbf{#1.}}

\AtBeginDocument{%
  \providecommand\BibTeX{{%
    \normalfont B\kern-0.5em{\scshape i\kern-0.25em b}\kern-0.8em\TeX}}}

\makeatletter
\g@addto@macro\normalsize{%
  \abovedisplayskip 3pt plus1pt 
  \belowdisplayskip 3pt plus1pt
  \abovedisplayshortskip  0pt plus1pt%
  \belowdisplayshortskip  0pt plus1pt
}
\makeatother

\acrodef{CV}{computer vision}
\acrodef{IR}{information retrieval}
\acrodef{LLM}{large language model}
\acrodef{OOD}{out-of-distribution}
\acrodef{MDP}{Markov decision process}
\acrodef{NLP}{natural language processing}
\acrodef{NRM}{neural ranking model}
\acrodef{RL}{reinforcement learning}
\acrodef{RL-MARA}{Multi-grAnular Ranking Attack}
\acrodef{MoE}{mixture-of-experts}

\copyrightyear{2025}
\acmYear{2025}
\setcopyright{cc}
\setcctype{by}
\acmConference[SIGIR '25]{Proceedings of the 48th International ACM SIGIR Conference on Research and Development in Information Retrieval}{July 13--18, 2025}{Padua, Italy}
\acmBooktitle{Proceedings of the 48th International ACM SIGIR Conference on Research and Development in Information Retrieval (SIGIR '25), July 13--18, 2025, Padua, Italy}\acmDOI{10.1145/3726302.3730049}
\acmISBN{979-8-4007-1592-1/2025/07}

\makeatother


\author{Yu-An Liu}
\orcid{0000-0002-9125-5097}
\affiliation{
 \institution{CAS Key Lab of Network Data Science and Technology, ICT, CAS}
 \institution{University of Chinese Academy of Sciences}
 \city{Beijing}
 \country{China}
}
\email{liuyuan21b@ict.ac.cn}

\author{Ruqing Zhang}
\authornote{Jiafeng Guo and Ruqing Zhang are the corresponding authors.}
\orcid{0000-0003-4294-2541}
\affiliation{
 \institution{CAS Key Lab of Network Data Science and Technology, ICT, CAS}
 \institution{University of Chinese Academy of Sciences}
\city{Beijing}
 \country{China}
}
\email{zhangruqing@ict.ac.cn}

\author{Jiafeng Guo}
\authornotemark[1]
\orcid{0000-0002-9509-8674}
\affiliation{
 \institution{CAS Key Lab of Network Data Science and Technology, ICT, CAS}
 \institution{University of Chinese Academy of Sciences}
   \city{Beijing}
 \country{China}
}
\email{guojiafeng@ict.ac.cn}

\author{Maarten de Rijke}
\orcid{0000-0002-1086-0202}
\affiliation{
 \institution{University of Amsterdam}
 \city{Amsterdam}
 \country{The Netherlands}
}
\email{m.derijke@uva.nl}

\author{Yixing Fan}
\orcid{0000-0003-4317-2702}
\affiliation{
 \institution{CAS Key Lab of Network Data Science and Technology, ICT, CAS}
 \institution{University of Chinese Academy of Sciences}
\city{Beijing}
 \country{China}
}
\email{fanyixing@ict.ac.cn}

\author{Xueqi Cheng}
\orcid{0000-0002-5201-8195}
\affiliation{
 \institution{CAS Key Lab of Network Data Science and Technology, ICT, CAS}
 \institution{University of Chinese Academy of Sciences}
   \city{Beijing}
 \country{China}
}
\email{cxq@ict.ac.cn}
\renewcommand{\shortauthors}{Yu-An Liu et al.}

\begin{document}

\title[On the Scaling of Robustness and Effectiveness in Dense Retrieval]{On the Scaling of Robustness and Effectiveness in Dense Retrieval}

\begin{abstract}
Robustness and Effectiveness are critical aspects of developing dense retrieval models for real-world applications. It is known that there is a trade-off between the two.
Recent work has addressed scaling laws of effectiveness in dense retrieval, revealing a power-law relationship between effectiveness and the size of models and data.
Does robustness follow scaling laws too?
If so, can scaling improve both robustness and effectiveness together, or do they remain locked in a trade-off?

To answer these questions, we conduct a comprehensive experimental study.
We find that:
\begin{enumerate*}[label=(\roman*)]
\item Robustness, including out-of-distribution and adversarial robustness, also follows a scaling law.
\item Robustness and effectiveness exhibit different scaling patterns, leading to significant resource costs when jointly improving both.
\end{enumerate*}
Given these findings, we shift to the third factor that affects model performance, namely the optimization strategy, beyond the model size and data size.
We find that: 
\begin{enumerate*}[label=(\roman*)]
\item By fitting different optimization strategies, the joint performance of robustness and effectiveness traces out a Pareto frontier. 
\item When the optimization strategy strays from Pareto efficiency, the joint performance scales in a sub-optimal direction. 
\item By adjusting the optimization weights to fit the Pareto efficiency, we can achieve Pareto training, where the scaling of joint performance becomes most efficient. 
\end{enumerate*}
Even without requiring additional resources, Pareto training is comparable to the performance of scaling resources several times under optimization strategies that overly prioritize either robustness or effectiveness.
Finally, we demonstrate that our findings can help deploy dense retrieval models in real-world applications that scale efficiently and are balanced for robustness and effectiveness.

\end{abstract}

\maketitle

\section{Introduction}
Dense retrieval models have achieved state-of-the-art effectiveness, which reflects overall performance under \emph{normal conditions}.
But they inherit vulnerabilities commonly associated with general neural networks \cite{zhong2023poisoning,liu2023black,reimers2021curse,liu2025robustness}, making them inherently disadvantaged in terms of robustness in \emph{abnormal situations}, such as handling \ac{OOD} data or adversarial attacks \cite{wu2022neural,liu2024robust,petroni2021kilt,thakur2beir,liu2025attachain}.
Due to this limitation, dense retrieval models face a trade-off between robustness and effectiveness \cite{thakur2beir,petroni2021kilt,zhong2023poisoning,zhao2022dense,liu2024perturbation}.
Understanding the interplay between these two aspects is critical for 
ensuring reliable ranking performance across diverse practical scenarios.

\begin{figure*}[t]
    \centering
    \includegraphics[width=\linewidth]{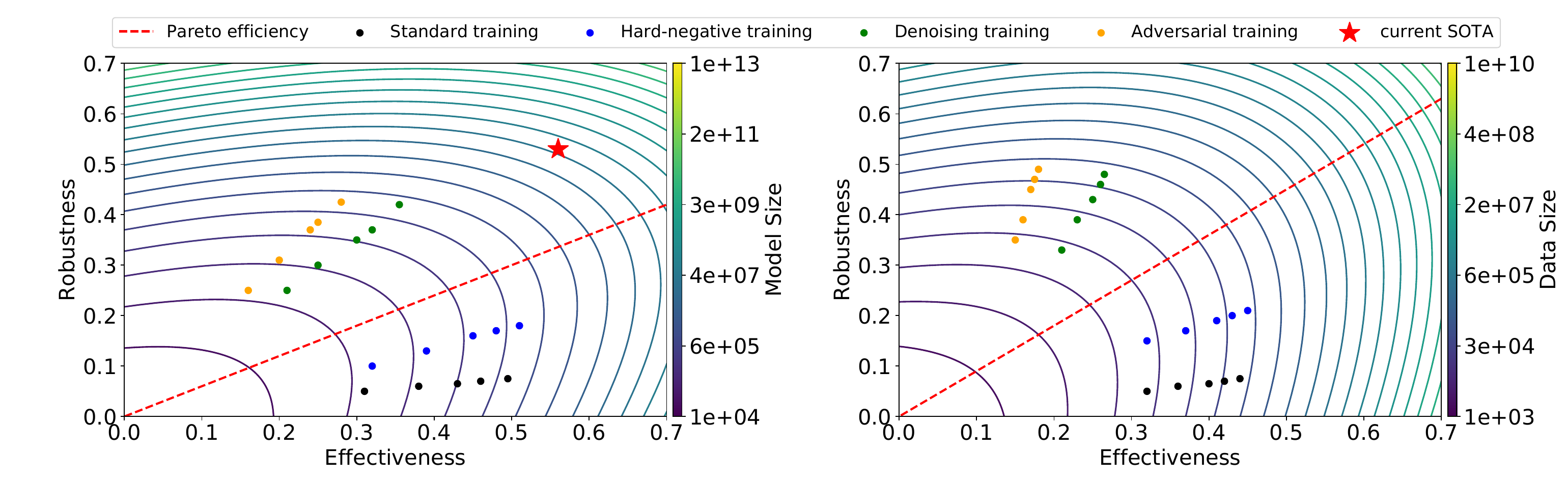}
    \caption{Joint scaling trends in robustness and effectiveness of the BERT model with respect to model size and data size across different optimization strategies. Effectiveness is evaluated on MS MARCO, and robustness is the average of the OOD robustness on BEIR and the adversarial robustness against ranking attacks on MS MARCO. The metric in the figure is obtained by inverse normalization of the (average) contrastive entropy. More experimental details are in Section \ref{Sec Joint scale}.}
    \label{robust-effect}
\end{figure*}

We explore robustness and effectiveness from the perspective of scaling, aiming to guide the development of dense retrieval models that excel in both dimensions.
Recent studies \cite{fang2024scaling,ni2022large} have examined scaling laws of effectiveness in dense retrieval, revealing that scaling model size and data volume enhance effectiveness and optimize the training process.
These findings motivate us to investigate whether scaling can simultaneously improve both robustness and effectiveness: \emph{Does robustness follow similar scaling laws?}

\heading{Scaling laws of robustness}
We address this question through experimental analysis, starting with an investigation of \emph{out-of-distribution (OOD) robustness} \cite{thakur2beir} in dense retrieval.
Our analysis focuses on the impact of two key factors, i.e., \emph{model size} and \emph{dataset size}, on robustness performance.
Following \cite{fang2024scaling}, we employ the contrastive entropy metric to assess the performance of dense retrieval models. 
Experiments are conducted using dense retrieval models implemented with various pre-trained language models, with non-embedding parameter sizes ranging from 0.5 to 87 million.
These models are trained on Chinese and English web search datasets, with training dataset sizes varying from 30K to 480K.
Our results show that robustness also adheres to a precise power-law scaling relationship w.r.t.\ both model size and dataset size.
We also observe that the scaling law for dataset size remain consistent across different annotation qualities. 
We investigate another critical aspect of robustness in IR scenarios \cite{castillo2011adversarial,wu2022neural,kurland2022competitive,liu2024robust,liu2025robust}: \emph{adversarial robustness}.
While scaling laws show variability (or ``noise'') in adversarial robustness for the original model, they are precise for dense retrieval models after adversarial training \cite{liu2024perturbation}.

\heading{Different scaling patterns between robustness and effectiveness}
With the scaling laws that govern robustness and effectiveness in dense retrieval, we aim to explore: \emph{Whether robustness and effectiveness can be improved together through scaling.}
The answer is: \emph{not exactly}.
The robustness of dense retrieval models is more sensitive to dataset size, while their effectiveness is more influenced by model size.
This re-introduces a trade-off between robustness and effectiveness in terms of resource requirements, as jointly improving requires substantial parameters and data costs.
E.g., our experiments show that achieving a 10\% improvement in both robustness and effectiveness for the state-of-the-art dense retrieval model (Llama2Vec, 7B) \cite{li-etal-2024-llama2vec} requires both a model size comparable to GPT-4 (175B) \cite{chatgpt} and scaling up training data 10-fold. 

Building on these findings, we consider a third critical factor influencing model performance, namely the optimization strategy, beyond just data and model.
By evaluating the performance of optimization strategies, our goal is to identify approaches that can jointly improve  robustness and effectiveness, even within constrained resource budgets, rather than relying solely on scaling up.
Our analysis reveals three key findings: 

\heading{Pareto efficiency exists in robustness and effectiveness} 
We keep the model and dataset size constant while exploring techniques such as hard-negative training \cite{xiongapproximate}, denoising training \cite{chen2023dealing}, and adversarial training \cite{liu2024perturbation} to evaluate their impact on both robustness and effectiveness.
As shown in Figure \ref{robust-effect}, fitting robustness and effectiveness under various optimization strategies reveals a Pareto frontier for joint performance. 
This represents \emph{Pareto efficiency}~\citep{watson-2013-strategy}, the red dashed line, where robustness cannot be better off without making effectiveness worse off, and vice versa.

\heading{Deviation from Pareto efficiency leads to suboptimal scaling} 
As shown in Figure \ref{robust-effect}, most representative training methods deviate from Pareto-efficient training by overemphasizing either robustness or effectiveness. 
Robustness and effectiveness are not appropriately assigned in these optimization objectives, leading to a sub-optimal scaling direction with inefficient performance improvement.
Take hard-negative training as an example: its overemphasis on model effectiveness on hard negatives limits the performance of robustness, resulting in attenuated gains in scaling model and data size.

\heading{Pareto training achieves balance and efficient scaling} 
To achieve Pareto efficiency between robustness and effectiveness, we propose a \emph{Pareto training} method.
Pareto training estimates the distance between the current model state and the Pareto efficiency at each training step, and adaptively adjusts the weights of robustness and effectiveness in the training objective. 
Compared with standard training, Pareto training can improve scaling efficiency by up to 2.5x within a certain range. 
Even without scaling, its joint performance is comparable to the gains achieved by simply scaling up, such as a multi-fold increase in dataset volume or in model size.

From a joint perspective of robustness and effectiveness, we show how Pareto efficiency can be used in practice to allocate resource budgets and highlighted the resource-friendliness of Pareto training. 
We hope our findings provide valuable insights for developing dense retrieval models that are not only effective but also robust.



\vspace*{-2mm}
\section{Problem Statement}
%
\subsection{Task Description}
\textbf{Dense retrieval.}
Dense retrieval serves as an implementation of the first-stage retrieval. 
Given a corpus and a query, the goal of the first-stage retrieval is to return a ranked list of top-$K$ most relevant documents based on the relevance score of each document in the corpus to the query  \cite{guo2022semantic,zhao2022dense,Karpukhin2020DensePR,xia2015learning}.
In dense retrieval, documents and queries are encoded into dense vectors; relevance is computed using these representations \cite{zhao2022dense}. 
Given a query $q$ and document $d$, dense retrieval models compute the relevance score $\operatorname{Rel}(q,d)$ as
\begin{equation}
\operatorname{Rel}(q,d) = f\left(\psi(q),\phi(d)\right),
\end{equation}
where $f(\cdot)$ is a interaction function typically realized by dot product, $\psi(\cdot)$ and $\phi(\cdot)$ are functions mapping queries and documents into $l$-dimensional vectors, respectively \cite{zhao2022dense}.
In this paper, we focus on the use of shared encoders for queries and documents, a widely adopted and effective approach in dense retrieval \cite{Karpukhin2020DensePR,khattab2020colbert,dong2022exploring}.

\heading{Effectiveness of dense retrieval} 
Effectiveness refers to the average ranking performance under conditions consistent with the training data.
Formally, given a dense retrieval model $f_{\mathcal{D}_\mathrm{train}}$ trained on the original training set $\mathcal{D}_\mathrm{train}$, its effectiveness $E$ refer to the ranking performance $\mathcal{R}_M$ under the original test data $\mathcal{D}_\mathrm{test}$:
\begin{equation}
    E = \mathcal{R}_M\left(f_{\mathcal{D}_\mathrm{train}} ; \mathcal{D}_\mathrm{test}\right).
\end{equation}

\heading{Robustness of dense retrieval}
In IR, robustness refers to the ability of a model to maintain ranking performance when facing unseen data from abnormal conditions.
In general, abnormal conditions include OOD queries and documents \cite{thakur2beir}, adversarial attacks \cite{wu2022prada}, and non-retrievable documents \cite{azzopardi2008retrievability}.
Given a well-trained dense retrieval model $f_{\mathcal{D}_\mathrm{train}}$, its robustness is derived from its ranking performance under unseen test data $\mathcal{D}^*_\mathrm{test}$:
\begin{equation}
\label{eq: robust}
    R = \mathcal{R}_M\left(f_{\mathcal{D}_\mathrm{train}} ; \mathcal{D}^*_\mathrm{test}\right).
\end{equation}
We focus on OOD robustness and adversarial robustness.
We leave the exploration of other types of robustness for future work.

\subsection{Training Setting}
\textbf{Model architecture.}
With the development of large-scale pre-trained language models, dense retrieval models have advanced significantly.
These models mostly adopt transformer-based architectures \cite{NIPS2017_3f5ee243}.
We use the BERT \cite{devlin2018bert} and ERNIE \cite{sun2019ernie} families of models as our base architecture. 
For English benchmarks, we use 24 BERT checkpoints from Google’s original release, ranging from BERT-Tiny (0.5 million parameters) to BERT-Base (82 million parameters). 
For Chinese benchmarks, we adopt the ERNIE series, which shares similar pre-training tasks with BERT and is trained on Chinese corpora.
To investigate the impact of model size on robustness, we experiment with BERTs of various sizes.


Following \cite{fang2024scaling}, we initialize these models' pre-trained language models followed by fine-tuning them on annotated datasets.
The output vector is typically extracted either from the [CLS] token representation or via mean pooling over the final transformer layer’s outputs.
To ensure comparability, a projection layer is added to each model to standardize embedding dimensions to 768.

\heading{Training data}
Following \cite{fang2024scaling}, we use two large-scale datasets to train our dense retrieval models.
\begin{enumerate*}[label=(\roman*)]
\item MS MARCO Passage Ranking \cite{nguyen2016ms} is an English web search dataset with about 8.8 million passages from web pages and 0.5 million training queries.
\item T2Ranking \cite{xie2023t2ranking} is a Chinese web search dataset with about 300k queries and over 2 million passages collected from real-world search engines. 
\end{enumerate*}
We take each query-positive-passage pair in the training data as an independent data point.
To investigate the impact of data size on robustness, we experiment with various numbers of training data points.
We randomly sample training data for each dataset from 30K to 480K, resulting in five sets of training data.

\heading{Evaluation data}
For effectiveness, we use the dataset on which the model was trained to evaluate it.
For OOD robustness, we use two benchmarks to measure the OOD robustness of the English and Chinese dense retrieval models, respectively.
For adversarial robustness, the test dataset consists of adversarial samples (highly ranked documents) generated by attacking the dense retrieval model.

For OOD robustness, we adopt benchmarks spanning diverse domains, with both broad and specialized topics, varying text types, sizes, query lengths, and document lengths.
The diversity of data corresponds to the OOD robustness challenges of reality.
\begin{enumerate*}[label=(\roman*)]
\item For English, we adopt the BEIR benchmark \cite{thakur2beir}, which includes 18 retrieval datasets with 9 different retrieval tasks.
\item For Chinese, there is no benchmark specific to OOD robustness, so we integrated 5 retrieval datasets from different domains to construct one, named BCIR.
\end{enumerate*}
They are e-commerce, entertainment video, and medical datasets in Multi-CPR \cite{long2022multi}, a medical community QA dataset, cMedQA2 \cite{zhang2018multi}, and a news retrieval dataset, TianGong-PDR \cite{wu2019investigating}.

\heading{Training method}
For the training method, we follow \cite{fang2024scaling} to adopt the most straightforward random negative sampling and in-batch negative techniques, as the \emph{standard training} method in Section \ref{sec:robust_laws}.
Given a query-passage pair $\left(q_i, d_i^{+}\right)$, we optimize the dense retrieval model with contrastive ranking loss:
\begin{equation}
\begin{split}
    &\mathcal{L}(\theta)={}\\
    &-\frac{1}{B} \sum_{i=1}^B \log \frac{\exp \left(\operatorname{Rel}\left(q_i, d_i^{+} ; \theta\right)\right)}{\exp \left(\operatorname{Rel}\left(q_i, d_i^{+} ; \theta\right)\right)+\sum_j \exp \left(\operatorname{Rel}\left(q_i, d_j^{-} ; \theta\right)\right)},
\end{split}    
\end{equation}
where $B$ is the training batch size, $\{d_j^{-}\}$ is the set of negative passages, and $\operatorname{Rel}\left(q, d ; \theta\right)$ denotes the relevance score evaluated by model $f$ with parameters $\theta$.
We fine-tune the models for a fixed 10,000 steps and randomly sample 256 negatives at each step.

\subsection{Evaluation Protocol}
\textbf{Evaluation for effectiveness.}
To evaluate the effectiveness of dense retrieval models, we follow \cite{fang2024scaling} to employ contrastive entropy as our evaluation metric $M$, which is easy to observe trends and has been shown to perform consistently with discrete ranking metrics like NDCG@K or MRR@K.
For each query-passage pair in the test set, the \emph{contrastive entropy} is calculated as:
\begin{equation}
\begin{split}
    &\operatorname{CE}(q_i, d_i^{+} ; \theta) = {}\\
    &-\log \frac{\exp \left(\operatorname{Rel}\left(q_i, d_i^{+} ; \theta\right)\right)}{\exp \left(\operatorname{Rel}\left(q_i, d_i^{+} ; \theta\right)\right)+\sum_j \exp \left(\operatorname{Rel}\left(q_i, d_j^{-} ; \theta\right)\right)},
\end{split}    
\end{equation}
where $\{d_j^{-}\}$ is the randomly selected negative set with 256 passages.

\heading{Evaluation for robustness}
To simulate unseen data, we adopt zero-shot evaluation on the robustness benchmarks, using the average contrastive entropy as a measure.
Given a benchmark with $n$ test datasets, the \emph{average contrastive entropy} across the benchmark is calculated as:
\begin{equation}
    \operatorname{ACE}(\theta) =  \frac{\sum_{i=1}^{n} \operatorname{CE}(\mathcal{D}_i;\theta)}{n},
\end{equation}
where $\mathcal{D}_i$ is a test dataset in the benchmark.
\begin{figure*}[t]
    \centering
    \includegraphics[width=\linewidth]{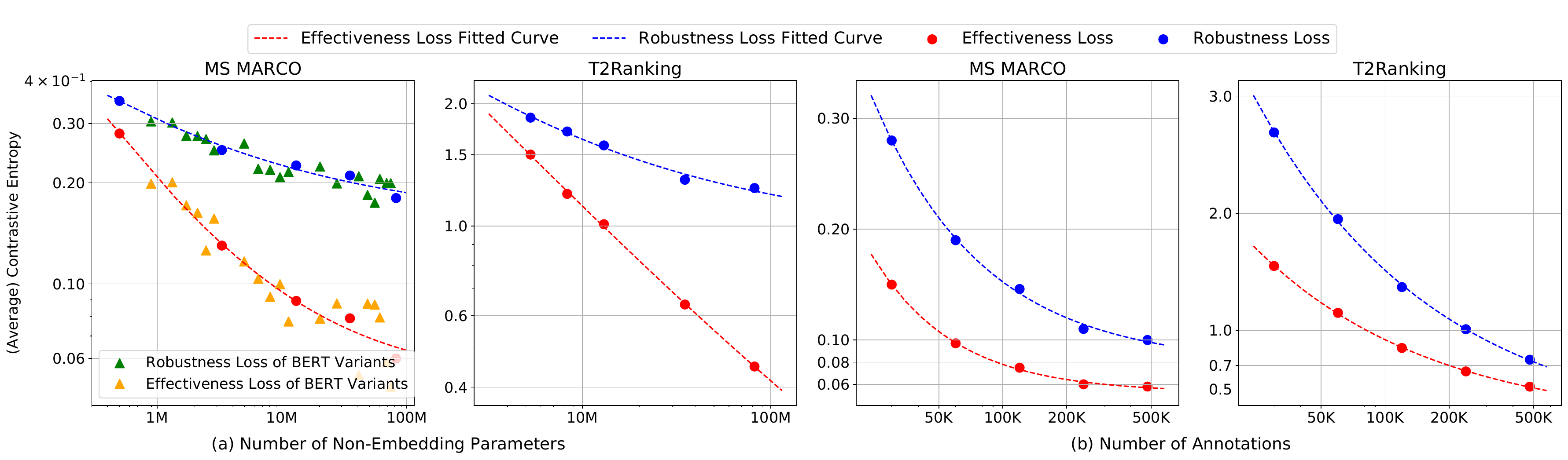}
    \caption{Scaling law for OOD robustness and effectiveness with model size (a) and data size (b) on MS MARCO and T2Ranking, respectively. Points represent the actual performance. Note that a decrease in loss represents an increase in performance.}
    \label{four_plots}
    
\end{figure*}

\section{Scaling Laws of Robustness}\label{sec:robust_laws}
In this section, we investigate the scaling laws of OOD robustness with respect to model size and data size, examine the impact of annotation quality on these scaling laws, and validate the identified scaling laws in the context of adversarial robustness.

\subsection{OOD Robustness w.r.t.\ Model Size}
\textbf{Experimental setup.}
We fine-tune models of different sizes using the complete training set. 
In line with \cite{fang2024scaling}, to prevent underfitting or overfitting, we avoid early stopping and instead report the best results obtained on the training dataset.

\heading{Scaling law w.r.t.\ model size}
The (average) contrastive entropy variation with model size is shown in Figure \ref{four_plots} (a). 
We see that the OOD robustness increases with the size of the model parameters.
On the MS MACRO dataset, square points represent the official checkpoints of 19 differently sized BERT models.
Based on our observations, we fit the scaling law of OOD robustness in terms of model sizes with log-linear functions following \cite{fang2024scaling,kaplan2020scaling}:
\begin{equation}\label{equ: model size}
    L(f) =  \left(\frac{M}{|f|}\right)^\mu + \delta_f,
\end{equation}
where $|f|$ represents the number of non-embedding parameters of the model, and $L(f)$ denotes the model’s contrastive entropy on the test set. The parameters $M$, $\mu$, and $\delta_f$ are the coefficients.

We employ the least squares method to fit the linear curve and obtain the parameters in Eq.~\ref{equ: model size} shown in Table \ref{tab:model size}.
The coefficient of determination ($R^2$) suggests the fitting error is acceptable. 

\begin{table}[t]
  \caption{Fitting parameters for model size scaling.}
  \label{tab:model size}
    \renewcommand{\arraystretch}{0.85}
   \setlength\tabcolsep{5pt}
  \begin{tabular}{l l cccc}
    \toprule
   Training dataset & Aspect & $M$ & $\mu$ & $\delta_f$ & $R^2$ \\
    \midrule
    \multirow{2}{*}{MS MARCO} & OOD & $3.70 \times 10^4$ & 0.55 & 0.05  & 0.997 \\
    & Effect. & $3.48 \times 10^4$ & 0.55 & 0.05  & 0.998 \\
    \midrule
    \multirow{2}{*}{T2Ranking} & Robust. & $4.23 \times 10^6$ & 0.46 & 0.96  & 0.989 \\
    & Effect. & $1.12 \times 10^7$ & 0.48 & 0.07  & 0.999 \\
  \bottomrule
\end{tabular}
\end{table}

\heading{OOD robustness scales smoothly with model size}
The results indicate that OOD robustness follows a precise power-law relationship with model size.
The scaling behavior of OOD robustness with model size remains consistent across our Chinese and English benchmarks.
When comparing the scaling laws of OOD robustness and effectiveness, we observe that effectiveness exhibits more dramatic variations with the number of model parameters, while OOD robustness changes more gradually. 
This suggests that larger models have a higher upper bound for ranking effectiveness. 
However, for OOD robustness, increasing the number of parameters may introduce more vulnerable neurons, potentially undermining the model's overall OOD robustness.

The coefficients $M$, $\mu$, and $\delta_f$ are derived from the dataset, where $\delta_f$ represents the inherent loss that cannot be optimized, possibly due to incorrect or incomplete annotations.
This observation also explains the higher $\delta_f$ values in the OOD robustness law, which can be attributed to the inconsistent annotation quality across datasets used in the OOD robustness benchmarks.

\subsection{OOD Robustness w.r.t.\ Data Size}
\textbf{Experimental setup.}
We fix the model size and use different sizes of training data to construct the training set.
Following \cite{fang2024scaling}, to avoid destabilizing effects of small-size models, we use model experiments at the largest scales in this experiment, i.e., BERT-Base and ERNIE-Base.

\heading{Scaling law w.r.t.\ data size}
The (average) contrastive entropy variation with data size is displayed in Figure \ref{four_plots} (b). 
OOD robustness improves as the data size increases. 
Based on the observation, we propose to fit the scaling law of OOD robustness in terms of data size as follows:
\begin{equation}\label{equ: data size}
    L(\mathcal{D}_\mathrm{train}) =  \left(\frac{D}{|\mathcal{D}_\mathrm{train}|}\right)^\eta + \delta_{\mathcal{D}_\mathrm{train}},
\end{equation}
where $|\mathcal{D}_\mathrm{train}|$ represents the number of annotated query-passage pairs, and $L(\mathcal{D}_\mathrm{train})$ denotes the model’s contrastive entropy on the test set. Parameters $D$, $\eta$, and $\delta_{\mathcal{D}_\mathrm{train}}$ are the coefficients.

We employ the least squares method to fit the linear curve and obtain the parameters in Eq.~\ref{equ: data size} shown in Table \ref{tab:data size}.
The coefficient of determination ($R^2$) suggests the fitting error is acceptable.

\begin{table}[t]
  \caption{Fitting parameters for data size scaling.}
  \label{tab:data size}
    \renewcommand{\arraystretch}{0.85}
   \setlength\tabcolsep{5pt}
  \begin{tabular}{l l cccc}
    \toprule
   Training dataset & Aspect & $D$ & $\eta$ & $\delta_{\mathcal{D}_\mathrm{train}}$ & $R^2$ \\
    \midrule
    \multirow{2}{*}{MS MARCO} & OOD & $4.34 \times 10^3$ & 0.83 & 0.08  & 0.998 \\
    & Effect. & $3.71 \times 10^3$ & 1.11 & 0.05  & 0.998 \\
    \midrule
    \multirow{2}{*}{T2Ranking} & Robust. & $1.93 \times 10^5$ & 0.51 & 0.12  & 0.996 \\
    & Effect. & $5.41 \times 10^4$ & 0.52 & 0.19  & 0.998 \\
  \bottomrule
\end{tabular}
\end{table}

\begin{figure}[h]
    \centering
    \includegraphics[width=\linewidth]{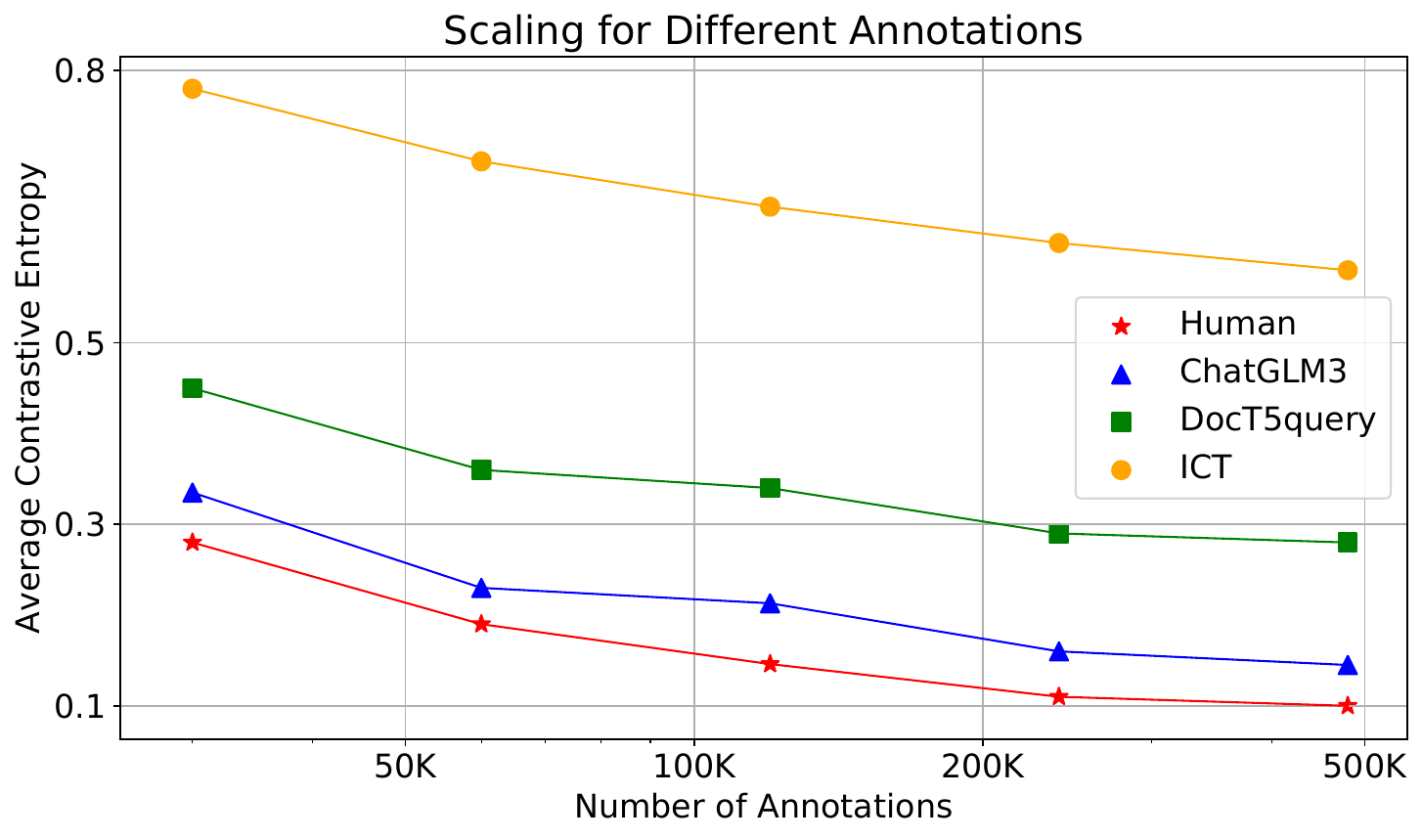}
    \caption{Scaling laws for OOD robustness with data size under different annotation methods.}
    \label{annotate}
\end{figure}

\heading{OOD robustness scales dramatically with data size}
The results indicate that OOD robustness follows a power-law scaling relationship with data size.
The OOD robustness performance with increasing data size shows similar trends across our Chinese and English benchmarks.
When comparing the scaling laws of OOD robustness and effectiveness, we observe that increasing the data size contributes more significantly to improving OOD robustness than effectiveness. 
This is likely because larger datasets provide a richer variety of samples, enabling models to better optimize their empirical decision boundaries. 
Establishing stable decision boundaries by exposing the model to sufficiently diverse data plays a crucial role in enhancing OOD robustness. 

The coefficients $D$, $\eta$, and $\delta_{\mathcal{D}\mathrm{train}}$ are derived from the dataset, where $\delta_{\mathcal{D}\mathrm{train}}$ represents the inherent loss associated with the dataset.
By comparing $\delta_f$ and $\delta_{\mathcal{D}\mathrm{train}}$, we find that the inherent losses from both the model and data perspectives are nearly identical. This observation further underscores the precision of the OOD robustness scaling law we have developed.

\vspace*{-3mm}
 \subsection{OOD Robustness for Annotation Quality}
Furthermore, we investigate whether the scaling effect for data size (Eq. \ref{equ: data size}) remains consistent across datasets of varying quality. 

\heading{Experimental setup}
To investigate the impact of different annotation qualities on robustness, we employ query generation techniques \cite{nogueira2019document} to create three distinct types of annotations.
\begin{enumerate*}[label=(\roman*)]
\item Inverse Cloze Task (ICT) \cite{lee2019latent} extracts key sentences from passages as the pseudo-query for the passage.
\item DocT5query \cite{nogueira2019document} uses a supervised generation model trained on human annotations to produce multiple queries for each passage.
\item ChatGLM3 \cite{zengglm} is a large language model (LLM) which generates relevant queries for given passages.
\end{enumerate*}
Following \cite{fang2024scaling}, for ICT and ChatGLM3, we generate a query for each positive document annotated by humans and for docT5query, we randomly sample 500,000 passages from the corpus for query generation, as the training data.
We take the OOD robustness performance in the English benchmark as an example.

\heading{Scaling law w.r.t.\ data size holds across different annotation qualities}
The results, shown in Figure \ref{annotate}, focus on the robustness of the English benchmark, with similar observations in the Chinese dataset.
When manually labeled data is replaced with data annotated using different methods, the pattern of OOD robustness scaling remains consistent. 
Models trained with manually labeled data achieve the highest OOD robustness, suggesting that manual annotation is still the most effective approach. 
This finding aligns with the observations of \cite{fang2024scaling} on effectiveness.

Interestingly, data annotated by generative models shows OOD robustness performance comparable to manual annotation, indicating that this method could be a cost-effective alternative for achieving satisfactory OOD robustness.
This highlights the potential for exploring hybrid annotation strategies, which may reduce annotation costs while maintaining or even enhancing the OOD robustness of the trained models.

\vspace*{-2mm}
\subsection{Extension to Adversarial Robustness} \label{Adversarial law}
To validate the reliability of the scaling laws for OOD robustness with respect to model size (Eq.~\ref{equ: model size}) and data size (Eq. \ref{equ: data size}), we test them in an adversarial robustness scenario.

\heading{Experimental setup}
We use a representative method for attacking IR models, the word substitution ranking attack (WSRA).
WSRA promotes a target document in rankings by replacing important words with synonyms \cite{wu2022prada}. 
We randomly sample 1,000 test queries in BEIR, for each sampled query, we randomly sample 1 document from 9 ranges in the candidates following \cite{wu2022prada,liu2024perturbation}, i.e., $[100,200], \ldots, [900,1000]$, respectively. 
We attack these 9 target documents to achieve their corresponding adversarial examples using WSRA. 
Finally, we evaluate the average contrastive entropy of dense retrieval models under the attacked list with 9 adversarial examples and its query as the adversarial robustness. 

In addition, we explore whether the model after adversarial training applies the scaling laws and robustness.
We use the adversarial training method proposed by \cite{liu2024perturbation} to train our model and observe the trend of their adversarial robustness.


\begin{figure*}[t]
    \centering
    \includegraphics[width=\linewidth]{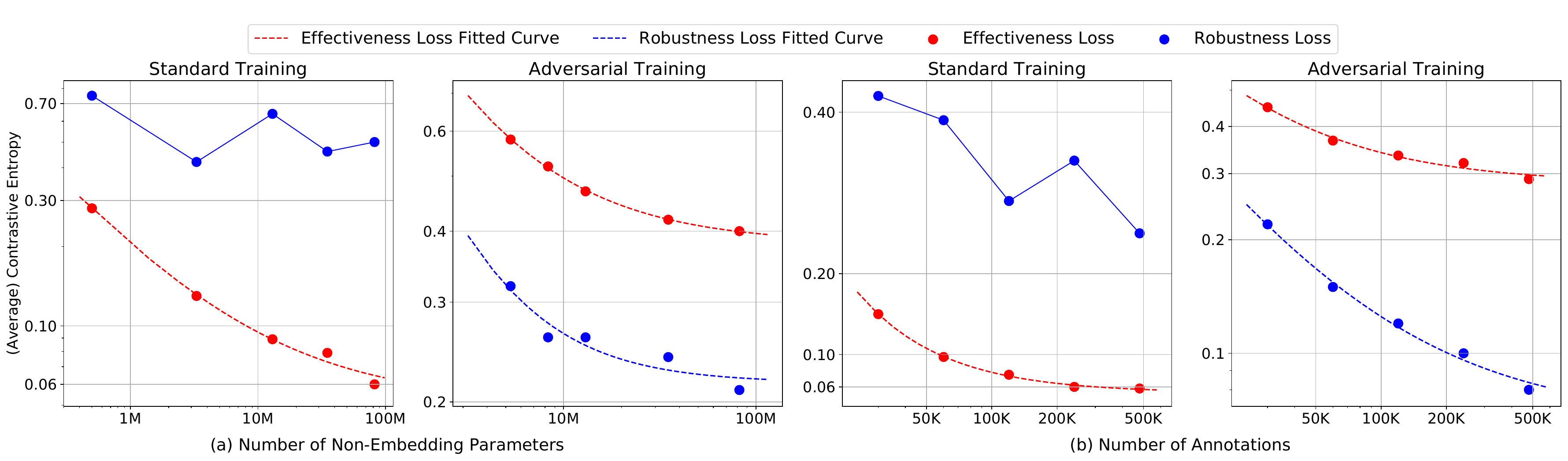}
    \caption{Scaling law for adversarial robustness and effectiveness with model size (a) and data size (b) on MS MARCO, respectively. We focus on adversarial training and standard training. The solid line indicates that it cannot be fitted into a scaling law.}
    \label{Fig: adversarial}
    
\end{figure*}

\heading{Adversarial robustness of the vanilla model is poorly suited for fitting}
The results of scaling laws for adversarial robustness w.r.t.\ data and model sizes for vanilla dense retrieval models are shown in Figure~\ref{Fig: adversarial}.
We take the robustness performance in the English benchmark as an example, with similar observations for the Chinese dataset.
Clearly, adversarial robustness poses greater challenges for dense retrieval models than OOD robustness.
While the general trend of adversarial robustness mirrors that of OOD robustness, it exhibits significant fluctuations and noise.
This can be attributed to the fact that adversarial samples are carefully crafted to exploit the model's vulnerabilities, making them particularly difficult to handle.
As a result, models often struggle with adversarial attacks, leading to relatively inconsistent performance.
When models have not been exposed to adversarial attacks, their adversarial robustness cannot be accurately captured by a scaling law.

\heading{Adversarial robustness of the model trained with adversarial examples is easily fitted}
When we examine the adversarial robustness of the dense retrieval model after adversarial training, as shown in Figure~\ref{Fig: adversarial}, adversarial training significantly enhances robustness across all model and data sizes.
Surprisingly, the robustness scaling for the adversarially-trained model becomes smooth and follows clear patterns.
The effectiveness and adversarial robustness of the adversarially-trained model fit the scaling law well, as shown in Table~\ref{tab:adversarial}.
This may be because the carefully crafted adversarial samples are entirely unfamiliar to the model, and after encountering similar examples, the model learns to behave more consistently, revealing the underlying scaling laws.
This finding aligns with the view that adversarial robustness is a form of extreme OOD robustness \cite{gokhale2022generalized,zou2024adversarial}.
The results suggest a general scaling law for robustness, paving the way for extending our findings to a broader and more diverse range of robustness scenarios.

\begin{table}[t]
  \caption{Fitting parameters for model and data size scaling on MS MARCO with adversarial robustness and effectiveness.}
  \label{tab:adversarial}
    \renewcommand{\arraystretch}{0.85}
   \setlength\tabcolsep{7pt}
  \begin{tabular}{l l cccc}
    \toprule
    & Aspect & $M$ & $\mu$ & $\delta_f$ & $R^2$ \\
    \midrule
    \multirow{2}{*}{Model size} & Adv. & $6.94 \times 10^5$ & 1.14 & 0.22  & 0.908 \\
     & Effect. & $8.34 \times 10^5$ & 0.87 & 0.38  & 0.999 \\
    \midrule
    & Aspect & $D$ & $\eta$ & $\delta_{\mathcal{D}_\mathrm{train}}$ & $R^2$ \\
    \midrule
    \multirow{2}{*}{Data size} & Adv. & $2.81 \times 10^3$ & 0.80 & 0.07  & 0.995 \\
     & Effect. & $3.74 \times 10^3$ & 0.87 & 0.28  & 0.986 \\
  \bottomrule
\end{tabular}
\end{table}

\vspace*{-2mm}
\subsection{Resource Budget Concerns}
From our results, it is clear that there are similar scaling laws for both robustness and effectiveness.
The key difference lies in their sensitivity to changes: robustness is more affected by variations in data, while effectiveness shows greater responsiveness to model size.
This suggests that improving both robustness and effectiveness simultaneously comes with a significant resource overhead.
Achieving substantial gains in both aspects would require a large investment in both model parameters and training data.
E.g., based on our scaling laws, further improving both effectiveness and robustness by 10\% from the current state-of-the-art model (Llama2Vec, 7B) \cite{li-etal-2024-llama2vec} would necessitate a model size on par with GPT-4 (175B) \cite{chatgpt} and scaling the training data by a factor of ten.
Both demands seem prohibitively expensive and practically challenging to meet.

So far, the scaling performance of robustness and effectiveness appears to present a trade-off: within limited resource budgets, it seems feasible to make significant gains in only one aspect.
This motivates us to investigate efficient approaches that could help improve the trade-off, potentially even optimizing both robustness and effectiveness jointly, rather than solely relying on scaling up.

\begin{figure*}[t]
    \centering
    \includegraphics[width=\linewidth]{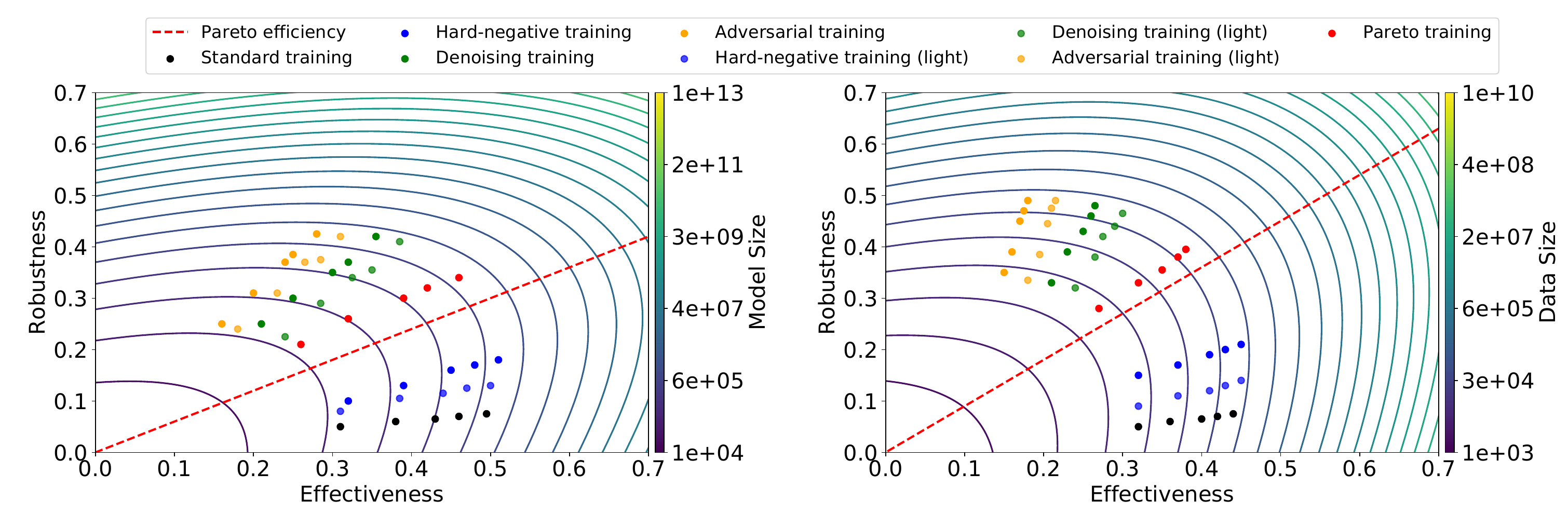}
    \caption{Joint scaling trends in robustness and effectiveness of the BERT model with respect to model size and data size across different optimization strategies. Effectiveness is evaluated on MS MARCO, and robustness is the average of the OOD robustness on BEIR and the adversarial robustness against ranking attacks on MS MARCO. For visualization purposes, the metric in the figure is obtained by inverse normalization of the (average) contrastive entropy.}
    \label{Fig joint scaling}
\end{figure*}

\vspace*{-1mm}
\section{Joint Scaling of Robustness and Effectiveness} \label{Sec Joint scale}

In this section, we conduct an in-depth exploration of the scaling of robustness and effectiveness through their joint performance.
We
introduce Pareto efficiency between robustness and effectiveness, and propose a Pareto training method that uses Pareto efficiency to jointly improve robustness and effectiveness.
We illustrate our findings using the performance of the BERT series on the English benchmark, with similar observations for the Chinese benchmark.

\vspace{-2mm}
\subsection{Pareto Efficiency}
The performance of dense retrieval is determined by three key factors: the model, the data, and the optimization strategy.
To efficiently improve the joint performance in terms of robustness and effectiveness, we explore optimization strategies and examine their impact across different models and data sizes.

\heading{Experimental setup}
Besides standard training, which uses random negative sampling and in-batch negative techniques, we adopt three optimization strategies that consider robustness in training dense retrieval models:
\begin{enumerate*}[label=(\roman*)]
\item hard-negative training \cite{xiongapproximate} uses BM25 \cite{StephenRobertson1994SomeSE} to mine hard negative samples, which are mixed with random negative samples during training;
\item denoising training \cite{chen2023dealing} estimates noise-invariant relevance by simulating noisy documents in the training process; and
\item adversarial training \cite{liu2024perturbation} enhances model robustness by modeling the boundary error between natural ranking and the ranking under adversarial perturbations.
\end{enumerate*}

For each strategy, we keep the overall data size constant and implement two variations: we keep 50\% and 75\% (balanced and light) of the original training data adopted by standard training to balance the training weights, respectively.
Robustness is measured using the average contrastive entropy, which includes two components: OOD robustness on BEIR and adversarial robustness on the original MS MARCO dataset.
For adversarial robustness evaluation data, we adopt the approach described in Section \ref{Adversarial law}; we randomly sample 1,000 queries, and for each query, we select 9 documents at different ranking intervals to generate adversarial samples.

\heading{Scaling performance varies from optimization strategies}
As shown in Figure \ref{Fig joint scaling}, we fit the robustness and effectiveness performance of various optimization strategies using model size and data size as contours, respectively.
Robustness and effectiveness both tend to improve with increases in model size and data size. Effectiveness is more strongly influenced by model size, while robustness is enhanced by data size.
This can be explained by differences in the scaling laws between the two as discussed in Section \ref{sec:robust_laws}.
Different optimization strategies follow similar scaling laws.

When comparing optimization strategies along a single contour:
\begin{enumerate*}[label=(\roman*)]
\item Standard training exhibits the worst robustness and limited effectiveness gains.
This suggests that focusing solely on effectiveness can lead to overfitting, and a lack of understanding of robustness can, in turn, constrain effectiveness.

\item Hard-negative training performs better than standard training in both robustness and effectiveness.
It introduces challenging and negative cases for comparison in the optimization objective, which improves the model’s ability to discriminate relevant cases, thereby enhancing both robustness and effectiveness.

\item Denoising training achieves improved robustness compared to hard-negative training, but at the cost of reduced effectiveness.
Denoising training enhances the model's robustness to irrelevant features by introducing random noise, but it also harms the relevant features of positive samples.

\item Adversarial training achieves the best robustness but performs poorly in terms of effectiveness.
An excessive focus on robustness can  compromise effectiveness.
\end{enumerate*}

\heading{Pareto efficiency exists between robustness and effectiveness}
By keeping the model size and data size constant, we can observe the trade-offs between robustness and effectiveness through different optimization strategies:
\begin{enumerate*}[label=(\roman*)]
\item at the ends of the contour lines, both robustness and effectiveness are relatively low;
\item as the optimization strategy shifts from focusing on only one aspect, both robustness and effectiveness increase; and 
\item as the weights of the optimization strategy approach balance, robustness and effectiveness exhibit Pareto efficiency.
\end{enumerate*}
At the Pareto efficiency, robustness has reached a proper position and cannot be better off without making effectiveness worse off, and vice versa.

\heading{Deviation from Pareto efficiency leads to suboptimal scaling}
When analyzing specific weighting ratios between effectiveness and robustness (e.g., the proportion of adversarial samples in adversarial training or the noise ratio in denoising training), we observe the following (see Figure \ref{Fig joint scaling}).
\begin{enumerate*}[label=(\roman*)]
\item The allocation of weights impacts both the model's current performance and its scaling behavior.
\item Examining the contours, we find that existing optimization strategies deviate from Pareto efficiency, resulting in suboptimal scaling directions.
While Pareto efficiency allows for the most efficient scaling, even at the Pareto point, the joint performance may match the equivalent scaling performance without further scaling.
\end{enumerate*}

Identifying Pareto efficiency is crucial for achieving an optimal balance between robustness and effectiveness. It provides a pathway to enhance joint performance while enabling efficient scaling.

\begin{figure*}[t]
    \centering
    \includegraphics[width=0.95\linewidth]{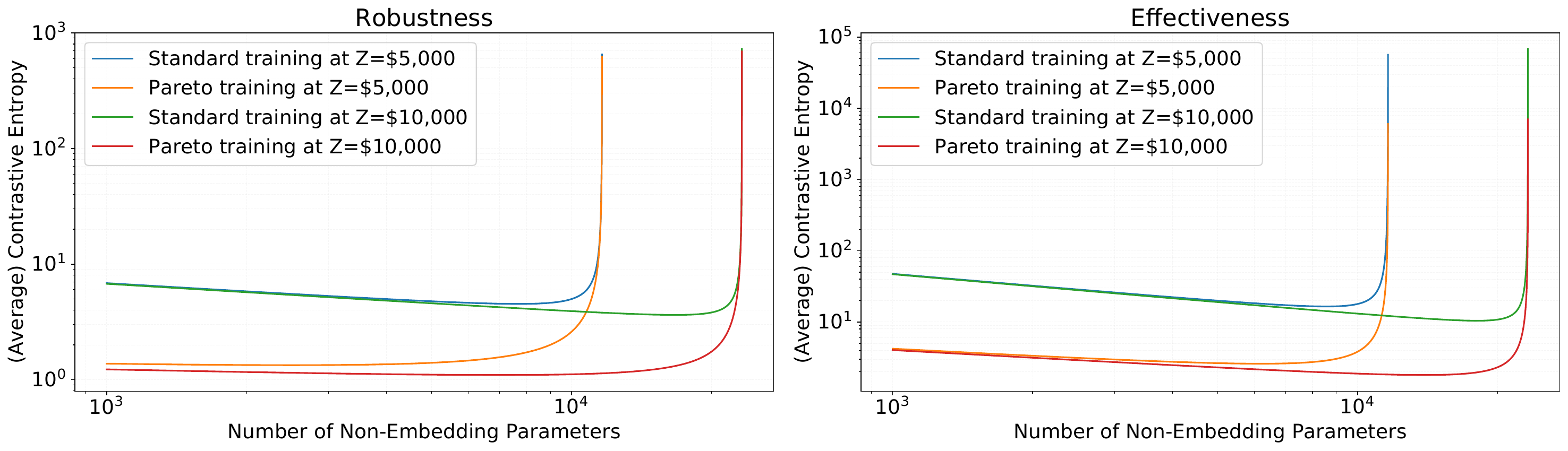}
    \caption{Predicted contrastive entropy of robustness and effectiveness for standard and Pareto training under limited budgets.}
    \label{Fig cost}
\end{figure*}

\vspace{-2mm}
\subsection{Pareto Training}
Building on the concept of Pareto efficiency, we propose Pareto training to develop dense retrieval models that balance robustness and effectiveness while enabling efficient scaling. 

\heading{Training method}
The key idea of the proposed Pareto training method is to dynamically adjust the weights of optimization objectives during training, allowing the joint performance of robustness and effectiveness to progressively approach Pareto efficiency.
A major challenge lies in the fact that the training losses for robustness and effectiveness do not always align with the model's final performance, making it difficult to achieve Pareto efficiency during training.
To address this, we adopt the concept of \emph{distributionally robust optimization}~\cite{goh2010distributionally} by dynamically adjusting the loss weights to estimate the gap with Pareto efficiency, thus deriving an approximate solution for the optimal weight $\omega$.

To initialize the weight $\omega_0$ we  calculate the ratio between robustness and effectiveness at Pareto efficiency.
We jointly optimize the robustness and effectiveness objectives using the following loss: 
\begin{equation}
    \mathcal{L}_\mathrm{Pareto}(\theta) = \omega \ell_R(\theta) + (1-\omega) \ell_E(\theta),
\end{equation}
where $\theta$ represents the parameters of the dense retrieval model, and $\ell_R(\cdot)$ and $\ell_E(\cdot)$ denote the robustness loss and effectiveness loss, respectively. 
The training loss on both robustness data and effectiveness data is implemented by pairwise loss functions.

At each training step $t+1$ ($t \geq 0$), we dynamically update the weight $\omega$ using the formula: 
\begin{equation}
    \omega^{t+1} = \omega^{t} - \eta\left(\frac{\ell^t_E(\theta)}{\ell^t_R(\theta)}-\frac{1}{\omega_0}\right),
\end{equation}
\begin{equation}
    \omega^{t+1} = \min(\max(\omega^{t+1},0),1),
\end{equation}
where $\eta$ is the learning rate and set to 0.1.

\heading{Experimental setup}
Pareto training can be adapted to various optimization strategies. 
We use adversarial training as an example, as it is somewhat distant from Pareto efficiency. 
We follow the adversarial training method from Section \ref{Adversarial law} and apply Pareto training to dynamically adjust training weights between adversarial sample optimization and original sample optimization to approach Pareto efficiency. 
We conduct experiments within the existing range of model and data sizes and observe the joint performance and scaling behavior on English benchmarks. 
For robustness testing, we continue to combine OOD robustness and adversarial robustness.

\heading{Pareto training efficiently optimizes robustness and effectiveness in a joint perspective}
In Figure \ref{Fig joint scaling} we observe that 
\begin{enumerate*}[label=(\roman*)]
\item Pareto training closely approaches Pareto efficiency in both model and data scaling. 
This indicates that our proposed training method can adaptively adjust the optimization direction, thereby improving the joint performance of robustness and effectiveness. 
\item Pareto training demonstrates a significant improvement in scaling efficiency, indirectly validating the accuracy of the Pareto efficiency we have fitted.
\item Even without scaling, the performance of the Pareto-trained model is equivalent to the result of large-scale scaling of the standard training model.
\end{enumerate*}

\vspace*{-1mm}
\section{Resource-Friendly Application}
We estimate the resource costs across the lifecycle of a dense retrieval model, including data preparation, model training, and model inference. 
We discuss, from the joint perspective of robustness and effectiveness, how to deploy resource-friendly dense retrieval models.

\vspace*{-1mm}
\subsection{Data-Model Joint Scaling Laws}
To study the resource costs, we consider the scaling laws of both data and models jointly.
Following \cite{fang2024scaling}, the joint effects of model size and data size can be approximated by:
\begin{equation}\label{equ: joint}
\begin{split}
L(f,\mathcal{D}_\mathrm{train}) = {} & L(f) + L(\mathcal{D}_\mathrm{train})\\
            {}\approx {}& \left[ \left( \frac{M}{|f|} \right)^{\frac{\mu}{\eta}} + \frac{D}{|\mathcal{D}_\mathrm{train}|} \right]^{\eta} + \delta,
\end{split}
\end{equation}
where $|f|$ and $|\mathcal{D}_\mathrm{train}|$ represent the model size and data size, respectively, and $M$, $D$, $\mu$, $\eta$, $\delta$ are coefficients.
This equation can be used to measure the total loss of either robustness or effectiveness.
In this section the robustness remains an average of OOD robustness and adversarial robustness; the setting is consistent with the Section \ref{Sec Joint scale}.
We take the English benchmark, approximate the parameters associated with standard training and Pareto training by numerical estimation; see Table \ref{tab:joint}.
Based on the joint scaling law, we can study the relationship between resource costs and performance.

\begin{table}[t]
  \caption{Estimated parameters in data-model joint scaling laws on English benchmark for standard and Pareto training.}
  \label{tab:joint}
    \renewcommand{\arraystretch}{0.85}
   \setlength\tabcolsep{4.5pt}
  \begin{tabular}{ll ccccc}
    \toprule
    Training & Aspect & $M$ & $D$ & $\mu$ & $\eta$ & $\delta$ \\
    \midrule
        \multirow{2}{*}{Standard} & Robust. & $2.11 \times 10^3$ & $2.99 \times 10^3$ & 0.10 & 0.78  & 0.01 \\
     & Effect. & $3.47 \times 10^4$ & $2.14 \times 10^3$ & 0.38 & 1.10  & 0.04 \\

    \midrule
    & Aspect & $M$ & $D$ & $\mu$ & $\eta$ & $\delta$ \\
    \midrule
        \multirow{2}{*}{Pareto} & Robust. & $1.96 \times 10^4$ & $2.57 \times 10^3$ & 0.25 & 0.80  & 0.02 \\
     & Effect. & $8.34 \times 10^5$ & $2.17 \times 10^3$ & 0.57 & 1.38  & 0.03 \\
  \bottomrule
\end{tabular}
\end{table}

\vspace*{-1mm}
\subsection{Resource Budget Allocation}
Approximately, the total cost of training and inference of a dense retrieval model with a data size of $|\mathcal{D}_\mathrm{train}|$ and model size $|f|$ is:
\begin{equation} \label{equ: cost}
Z(f,\mathcal{D}_\mathrm{train}) = Z_\mathrm{data} \cdot |\mathcal{D}_\mathrm{train}| + Z_\mathrm{train} \cdot |f| + Z_\mathrm{infer} \cdot |f|,
\end{equation}
\begin{equation}
Z_\mathrm{data} \approx 0.6, \; Z_\mathrm{train} \approx 3.22 \times 10^{-8}, \; Z_\mathrm{infer} \approx 0.43,
\end{equation}
where $Z_\mathrm{data}$, $Z_\mathrm{train}$, $Z_\mathrm{infer}$, represent cost factors corresponding to data preparation, training, and inference, respectively.
They are given by prior research \cite{fang2024scaling}, with the units in dollars.
Notably, in our setup, the data costs $Z_\mathrm{data}$ not only include manual annotation but also automatically generated data, such as adversarial examples. These costs are relatively lower than manual annotation, so we can treat them as having equivalent costs for the sake of simplicity.

By combining Eq.~\ref{equ: joint} and~\ref{equ: cost}, we can observe the trend of how robustness and effectiveness change with model size scaling under a fixed budget. 
We select two budget levels ($Z=\$5,000$ and $Z=\$10,000$) and compare the trends of standard training and Pareto training in terms of robustness and effectiveness. The results are shown in Figure \ref{Fig cost}.
\begin{enumerate*}[label=(\roman*)]
\item Within a certain range, both methods can achieve improvements in robustness and effectiveness by increasing the model size. 
However, indiscriminate scaling can lead to a limited available budget for data, which ultimately harms performance. 
\item Due to its relative insensitivity to model scaling, robustness often experiences performance degradation earlier. 
When allocating the budget, we need to  consider both robustness and effectiveness to achieve a balanced performance in dense retrieval models.
\item By comparison, due to its ability to achieve efficient scaling,
Pareto training exhibits a scaling efficiency improvement of about 2.5 times compared to standard training when the budget is 5,000.
It can fully use limited budgets, and scale to larger models to achieve better joint performance in terms of robustness and effectiveness.
\end{enumerate*}
\vspace*{-2mm}
\section{Related Work}

\textbf{Dense retrieval.} As the size of neural networks increases, dense vectors output by the models contain rich information and exhibit strong discriminative power, making them suitable for indexing and distinguishing documents \cite{Karpukhin2020DensePR,zhao2022dense,guo2022semantic}. 
This has led to the emergence of dense retrieval, which stands out among  retrieval model families due to its effectiveness \cite{lin2022proposed}. Subsequent research has focused on enhancing the effectiveness of dense retrieval, such as increasing model capacity \cite{ni2022large} and expanding training data \cite{craswell2021ms}.
Only optimizing for effectiveness may lead to a loss in robustness.

Robustness is another crucial metric for dense retrieval models. 
Existing studies have found that despite their remarkable effectiveness, dense retrieval models sometimes fall short in robustness compared to traditional sparse retrieval models \cite{thakur2beir,wang2022micro,wang2021discover}. 
Existing work mainly focuses on OOD robustness and adversarial robustness for dense retrieval.
E.g., \citet{thakur2beir} and \citet{liu2023robustness} have revealed flaws of dense retrieval models in terms of OOD robustness, while \citet{liu2023black} \citet{zhong2023poisoning} have identified vulnerabilities to adversarial robustness.
Concerns about robustness have hindered the widespread application of dense retrieval models in real-world scenarios, and current research lacks a precise understanding of the scaling laws of robustness.

Both robustness and effectiveness are important for dense retrieval models. 
Some work claims that robustness and effectiveness are conflicting goals that are difficult to optimize simultaneously \cite{tsipras2019robustness,wu2022neural}. 
Others have found that models with strong effectiveness also demonstrate good robustness, revealing that there is potential for the two to be co-improved \cite{liu2024perturbation, liu2024multi}. 
Nevertheless, existing work lacks quantitative analysis, leaving a limited understanding of their relationship. 
In this paper, we explicitly address the relationship between robustness and effectiveness by investigating their scaling laws and their trade-off performance.

\heading{Scaling laws} 
\citet{zipf2016human} reveals an inverse relationship between word frequency and its rank in the frequency distribution of natural language. 
Heaps' law \cite{gelbukh2001zipf} describes the growth pattern of vocabulary size with the total number of words in a document, becoming a key principle in estimating inverted indexes in information retrieval. 
With the development of neural networks, research on scaling laws has gradually shifted to the size of neural models, dataset sizes, and computational resources. E.g., \citet{hestness2017deep} discover a power-law relationship, which has since been extended to larger models \cite{kaplan2020scaling} and eventually quantified exactly \cite{hoffmann2022empirical}. These findings provide insights for predicting model performance and rationally allocating training resources.
In IR, GTR uses scaling laws to increase model size and enhance dense retrieval performance \cite{ni2022large}. \citet{fang2024scaling} precisely fit the scaling laws of dense retrieval effectiveness using formulas. These works primarily focus on the scaling laws of effectiveness, neglecting the changes in model robustness and the relationship between robustness and effectiveness.
\vspace*{-3mm}
\section{Conclusion}
We have presented a comprehensive analysis of scaling laws for robustness and effectiveness in dense retrieval, exploring methods to jointly optimize both while maintaining efficiency.
By scaling model and data sizes, we observe that the robustness of dense retrieval adheres to scaling laws. 
The scaling patterns of robustness and effectiveness differ, resulting in significant resource overheads for joint optimization.
We reveal the existence of Pareto efficiency between robustness and effectiveness, with typical optimization strategies often yielding suboptimal scaling performance due to deviations from this balance.
To address this, we propose Pareto training, which achieves an optimal trade-off between robustness and effectiveness, enabling efficient model scaling.
Even without scaling, Pareto training consistently enhances both robustness and effectiveness. 
We also show how Pareto efficiency can guide resource allocation for practical deployment scenarios.
We hope that our findings can contribute to the development of resource-friendlier IR.

\heading{Limitations and future work} 
\begin{enumerate*}[label=(\roman*)]
\item For OOD robustness, the BEIR benchmark is primarily used for evaluation. However, the uneven data distribution across datasets in the benchmark may impact the stability of our results. 
In future work, more realistic simulations of robustness scenarios warrant further investigation.
\item While this study focused on OOD robustness and adversarial robustness, which are well-established in information retrieval, future research should explore a broader range of robustness types, including performance variance \cite{zhang2013bias} and retrievability \cite{azzopardi2008retrievability}.
\item We investigate scaling laws for the robustness of dense retrieval models in terms of model size, data size, optimization strategy, and annotation quality, leaving aspects like scaling laws for computational costs \cite{kaplan2020scaling} and attack costs \cite{howe2024effects} for future work.
\item For model architecture, we focus on representative dual-encoder architectures that map queries and documents into embeddings of the same dimension through a shared encoder. Other architectures, such as multi-vector generation \cite{khattab2020colbert}, hybrid models \cite{shen2023unifier}, or interaction-based dense retrieval models \cite{humeaupoly}, may offer different scaling performances and deserve further exploration in future studies.
\item We examine three optimization strategies; methods such as customized pre-training \cite{ma2021prop}, distillation \cite{Qu2021RocketQAAO}, and domain-adaptive training \cite{ma2021zero} remain unexplored. These strategies could impact model robustness.
\item Due to time and resource constraints, this study focuses on scaling models within BERT-level pre-trained architectures, with future work planned for validation on large language models.
\end{enumerate*}

\vspace*{-1mm}
\begin{acks}
This work was funded by the National Natural Science Foundation of China (NSFC) under Grants No. 62472408, 62372431 and 62441229, the Strategic Priority Research Program of the CAS under Grants No. XDB0680102 and XDB0680301, the National Key Research and Development Program of China under Grants No. 2023YFA1011602, the Youth Innovation Promotion Association CAS under Grants No. 2021100, the Lenovo-CAS Joint Lab Youth Scientist Project, and the project under Grants No. JCKY2022130C039.  This work was also (partially) funded by the Dutch Research Council (NWO), under project numbers 024.004.022, NWA.1389.20.183, and KICH3.LTP.20.006, and the European Union’s Horizon Europe program under grant agreement No. 101070212.

All content represents the opinion of the authors,
which is not necessarily shared or endorsed by their respective employers and/or sponsors. 
\end{acks}

\bibliographystyle{ACM-Reference-Format}
\balance
\bibliography{references}

\end{document}